\documentclass[a4paper,useAMS,usenatbib]{mn2e}

\usepackage[pdftex]{graphicx}
\usepackage{amssymb,amsmath}
\usepackage{booktabs}
\usepackage{afterpage}
\usepackage{pdflscape}
\usepackage{multirow}

\usepackage{array}
\usepackage{hyperref}

\usepackage{capt-of}

\usepackage{lipsum}

\title[\bf Outer Galactic disk A and F stars]{\bf A study of full space motions of outer Galactic disk A and F stars in two deep pencil-beams}
\author[A. Harris et. al.]{{\parbox{\textwidth}{A. Harris$^{1}$\thanks{E-mail: a.harris7@herts.ac.uk}, J. E. Drew$^1$,  M. Mongui\'{o}$^{1}$}}
\\\\
$^{1}$School of Physics, Astronomy and Mathematics, University of Hertfordshire, College Lane, Hatfield AL10 9AB, UK }
\begin{document}

\date{}

\pagerange{\pageref{firstpage}--\pageref{lastpage}} \pubyear{2019}

\maketitle

\label{firstpage}

\begin{abstract}


A and F stars can be used as probes of outer Galactic disk kinematics: here we extend the work of \citet{Harris2018} by crossmatching their A/F sample with Gaia DR2 to bring in proper motions. These are combined with the already measured radial velocities and spectro-photometric distances to obtain full space motions. We use this sample of 1173 stars, located in two pencil-beam sightlines ($\ell=178^\circ$ and $\ell=118^\circ$), to sample the Galactocentric velocity field out to almost $R_G=15$\,kpc. We find there are significant differences in all three (radial, azimuthal and vertical) kinematic components between the two directions. The rotation curve is roughly flat in the anticentre direction, confirming and extending the result of \citet{Kawata2018} thanks to the greater reach of our spectro-photometric distance scale. However at $\ell=118^\circ$ the circular velocity rises outwards from $R_G=10.5$\,kpc and there is a more pronounced gradient in radial motion than is seen at $\ell=178^\circ$. Furthermore, the A star radial motion differs from the F stars by $\sim10$\,km\,s$^{-1}$. We discuss our findings in the context of perturbers potentially responsible for the trends, such as the central bar, spiral arms, the warp and external satellites. Our results at $\ell=178^\circ$ are broadly consistent with previous work on K giants in the anticentre, but the kinematics at $\ell=118^\circ$ in the Perseus region do not yet reconcile easily with bar or spiral arm perturbation. 

\end{abstract}

\begin{keywords}
Galaxy: disc -- Galaxy: kinematics and dynamics -- stars: early-type -- methods: observational
\end{keywords}

\setlength{\extrarowheight}{5pt}
\section{Introduction}
\label{sec:introduction}

Tracing the kinematics of the Galactic disk allows us to map out its structure and so gain insight into its formation and evolution. The disk is expected to be rich in substructure imprinted from dynamical processes such as resonant effects from the central bar \citep{Monari2014}, the spiral arms \citep{Monari2016, Grand2016}, and infalling and external satellites \citep{Gomez2013, Antoja2018}. The mean structure contains important information about the Galactic potential and yet remains uncertain. To form a complete picture it is vital to study not only the accessible inner disk and Solar Neighbourhood, but also the outer disk. So far, the outer disk is less well known.

With the DR2 release from the Gaia mission \citep{Brown2018, Prusti2016}, there is now an abundance of kinematic data in the form of positions, parallaxes, proper motions and some radial velocities. However for the faint distant stars that populate the outer disk, Gaia DR2 provides valuable proper motion data, but not radial velocities and parallaxes good enough for use in kinematic studies. Previous studies have tended to use either masers in star-forming regions \citep{Honma2012, Reid2014} or clump giants \citep{Lopez-Corredoira2014, Huang2016, Tian2017}, both boasting reasonably well-defined distances. But both of these tracers have down sides too - masers in star forming regions are sparse in the outer disk, making it difficult to sufficiently sample the kinematic structure, and clump giants are old and hence are subject to large kinematic scatter and asymmetric drift. More recently, \citet{Kawata2018} used $>10^6$ stars from Gaia DR2 to study the azimuthal and vertical velocity field out to Galactocentric radius $R_G<12$\,kpc in the anticentre direction, but the lack of radial velocity information and large parallax uncertainties for the fainter more distant stars prevented them from delving further into the outer disk \citep[see also][]{Katz2018}.

There is however another path to explore. \citet{Harris2018} (hereafter H18) showed that A and F stars can begin to be used as probes of outer Galactic disk kinematics on combining radial velocities and spectro-photometric distances. These stars offer the following advantages: i) they are intrinsically relatively luminous with absolute magnitudes in the $i$ band of $\sim$0 to 3, ii) as younger objects ($<1$\,Gyr), they have experienced significantly less scattering within the Galactic disk \citep{DehnenBinney1998}, iii) A stars especially are efficiently selected from photometric H$\alpha$ surveys (H18). Here we extend the work of H18 by completing the kinematics of their A and F star sample by bringing in the since-released Gaia DR2 proper motions. We examine the resultant Galactocentric radial, azimuthal and vertical velocity fields in their two pencil-beam sightlines in the outer disk at longitudes $\ell=118^\circ$ and $\ell=178^\circ$. We re-use the H18 spectro-photometric distance scale, which reaches out to Galactocentric radii of 14-15\,kpc.

The layout of the paper is as follows: in section \ref{sec:method} we describe the data, coordinate systems and how we compute full space motions. We also demonstrate the advantages of the H18 spectro-photometric distance scale relative to parallax-based alternatives. The results are presented in section \ref{sec:results}, detailing the profile of radial, azimuthal and vertical velocity with Galactocentric radius along the two sightlines. We also compare the kinematics and velocity ellipsoids of the A and F star populations. In section \ref{sec:disc1}, we discuss possible kinematic perturbers that may be at work and examine the results with these in mind. We end the paper with our conclusions; that there are departures from axisymmetry such that (to good precision) the disk circular speed rises with Galactocentric radius at $\ell=118^\circ$, whilst remaining flat at $\ell=178^\circ$. We also find a strong trend in Galactocentric radial velocity at $\ell=118^\circ$ that is much weaker at $\ell=178^\circ$.

\section{Sample and Method}
\label{sec:method}
\begin{figure}
 \centering
  \includegraphics[width=0.49\textwidth]{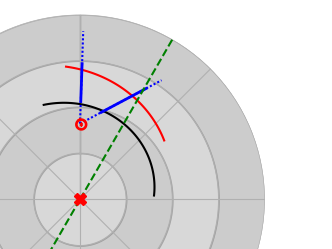}
 \caption{\it{A sketch of the Milky Way showing the locations of the sightlines, segments of the Perseus (black) and Outer (red) Arms \citep{Reid2014}, and an approximate (30$^\circ$) alignment of the central bar's major axis (green dashed). The Galaxy is rotating clockwise, with the centre shown by the red cross. The Sun is located at $R_0=8.2$\,kpc (red Sun symbol). Blue dashed lines show the full extent of the sightlines, and solid blue lines represent the actual distance range analysed. The $\ell=178^\circ$ sampling spans $10.7<R_G$ (kpc) $<14.7$, while the $l=118^\circ$ sightline captures $9.7<R_G$ (kpc) $<14.0$. The grey contours are spaced at distance intervals of $R_G=5$\,kpc. }}
\label{fig:sketch}
\end{figure}

The sample we use is that from H18\footnote{The data from H18 can be found here: \url{http://vizier.u-strasbg.fr/viz-bin/VizieR?-source=J/MNRAS/475/1680}}, comprising spectra of 1173 A and F stars. The sample was selected using the IPHAS\footnote{INT (Isaac Newton Telescope) Photometric H$\alpha$ Survey of the Northern Galactic Plane, see \citet{Drew2005}} $r-i$, $r-H\alpha$ colour-colour diagram, and spectra were gathered using MMT's multi-object spectrograph, HectoSpec. The stars have apparent magnitudes $15\lesssim G \lesssim19$ (equivalently $14\lesssim i \lesssim18$), sampling heliocentric distances of $2-10$\,kpc. They are located in two pencil-beams of $1^\circ$ diameter in the Galactic plane, at ($\ell$, $b$) = ($118^\circ$, $2^\circ$) and ($178^\circ$, $1^\circ$) (see figure \ref{fig:sketch}). Radial velocities and stellar parameters were measured using a MCMC-assisted parameter fitting routine relative to synthetic spectra as templates, which were calculated using the approach of \citet{Gebran2016} and \citet{Palacios2010}. Spectro-photometric distances were calculated taking into account measured extinctions and with the use of Padova isochrones \citep{Bressan2012, Chen2015}.  For a detailed description of the determination of radial velocities, stellar parameters and distances, see H18.

There are 780 stars in the $\ell=118^\circ$ sightline, and 393 in the $\ell=178^\circ$ sightline. We crossmatch this sample with the Gaia DR2 database, finding matches for all objects. This provides proper motions with typical uncertainties of $<0.2$\,mas\,yr$^{-1}$, or typical percentage uncertainties of $<15\%$. We then apply the following quality cuts:
\begin{itemize}
\item Following the suggestions of \citet{Lindegren2018}, we apply a quality cut depending on the unit weight error, $u_L$, of the Gaia data, which is a goodness-of-fit statistic on the model used to determine the astrometric parameters. We remove objects from the sample that have $u_L > 1.2 \times \max(1, \exp(-0.2(G-19.5)))$, where $G$ is the magnitude in the Gaia $G$ band. This quality cut results in 23 objects being removed from the $\ell=118^\circ$ sample, and 6 from the $\ell=178^\circ$ sample. 
\item Again following \citet{Lindegren2018}, we apply a cut based on the number of `visibility periods used' of the Gaia data, indicating an astronometrically well-observed source. We remove objects with $<8$ visibility periods.  This affects no objects at $\ell=118^\circ$. A further 2 objects are removed from the $\ell=178^\circ$ sample.
\item Finally, we remove objects with unrealistically large spectro-photometric distances. We set this limit to a heliocentric distance of 10\,kpc, since it is the maximum distance expected from the initial target selection. This cut removes a further 27 objects from the $\ell=118^\circ$ sample, and 10 from the $\ell=178^\circ$ sample. In our analysis we further reduce the heliocentric distance range considered to 6.5\,kpc ($\ell=178^\circ$) and 8.1\,kpc ($\ell=118^\circ$). 
\end{itemize}
The final sample size is then 730 stars at $\ell=118^\circ$ and 375 at $\ell=178^\circ$. By combining the now-complete kinematic data of our sample with our spectro-photometric distances, we have a full 6D phase-space solution.


\subsection{Distance scale}
\label{sec:dist}
For the magnitude range of the target stars ($15<G<19$), the measurement uncertainty on Gaia DR2 parallaxes range from $\sim0.04$\,mas for the brightest objects to $\sim0.3$\,mas for the faintest. The majority have an uncertainty around 0.1\,mas. Over the distance range targeted (2-10\,kpc, or a parallax range of 0.1-0.5\,mas), these uncertainties are significant, amounting to percentage uncertainties of $\sim20-100\%$. Additionally, the Gaia DR2 parallaxes carry systematic errors of order $\pm0.1$\,mas on global scales, with actual magnitude and distribution of the errors unknown \citep{Lindegren2018}. For these reasons, it is not obvious that Gaia DR2 parallax information would provide any improvement to our spectro-photometric distance scale. To check this, we have made comparisons with 2 alternative distance scales that make use of the Gaia DR2 parallaxes: i) \citet{Bailer-Jones2018} (hereafter BJ18) who infer a distance scale using a prior that varies according to a Galaxy model, and ii) the \textsc{topcat} `distanceEstimateEdsd' function (hereafter EDSD) which uses an exponentially decreasing space density prior for a chosen length scale (we use L=1500\,pc) to estimate the distance \citep{Luri2018}.

\begin{figure}
 \centering
  \includegraphics[width=0.49\textwidth]{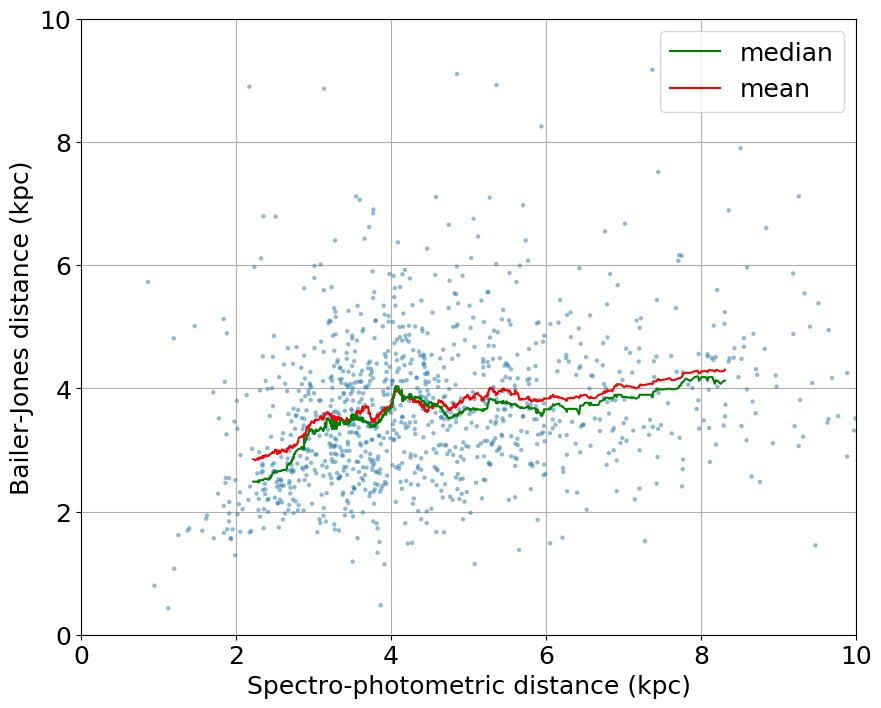}
 \caption{\it{A comparison of the HectoSpec spectro-photometric distances with the \citet{Bailer-Jones2018} distances. The red line is the running median and green line the running mean.}}
\label{fig:dist-comp}
 \end{figure}

Figure \ref{fig:dist-comp} compares the HectoSpec spectro-photometric distance scale with that from BJ18. The comparison with EDSD is not shown since it is very similar. We see that at $\geq4$\,kpc the BJ18/EDSD distances become significantly smaller than our spectro-photometric distances. This is signalling that at distances of $3-4$\,kpc the parallax measurements are becoming sufficiently imprecise that the prior in the parallax inversion takes over. Hence it is not appropriate to rely on Gaia DR2 parallax-based distance scales to probe further out than a few kpc from the Solar neighbourhood. \citet{Hogg2018} support this view,  instead making use of the APOGEE–Gaia–2MASS–WISE overlap to train a model to predict parallaxes from spectro-photometric data of red giant-branch stars out to heliocentric distances of 20\,kpc, without the use of a prior. Similarly, our spectro-photometric scale does not depend on a distance-related prior. 

In addition to this, the BJ18/EDSD distances suffer from large asymmetric uncertainties. The median percentage uncertainty for the BJ18 distances is $24\%$ for the inside tail of the probability distribution, and $39\%$ for the outside tail. Similarly for the EDSD distances, the median percentage uncertainty is $14\%$ for the inside tail and $59\%$ for the outside tail. The uncertainties on our spectro-photometric distances are symmetric and the percentage uncertainties range between $\sim5-25\%$, with a median of just $13\%$. Hence continued use of the spectro-photometric distance scale is warranted as it is free of the aforementioned systematic errors, benefits from relatively small, symmetric uncertainties, and most importantly can be trusted out to distances beyond those that can be inferred from Gaia DR2 parallaxes.

\subsection{Coordinate systems and conversion to the Galactocentric frame}

Throughout this paper we mainly use the Galactic coordinate system defined by longitude $\ell$ and latitude $b$, with heliocentric radial velocity $v_r$, defined as positive if the object is moving away from the Sun, and tangential velocities $v_\ell$, $v_b$, regarded as positive in the direction of increasing $\ell$ and $b$. The tangential velocities are derived from Gaia DR2 proper motions, which we convert from RA, DEC to $\ell$, $b$ following \citet{Poleski2018}, making use of the covariance matrix when propagating the errors.

We also use a cylindrical Galactocentric coordinate frame, defined by: Galactocentric azimuth $\phi$ measured from the centre-anticentre line with $\phi$ increasing in the direction of Galactic rotation; Galactocentric radius $R_G$, and the distance from the mid plane $Z$. In this frame, the velocities are ($u$, $v$, $w$), with $u$ being the Galactocentric radial velocity, positive in the direction towards the Galactic centre, $v$ being the azimuthal velocity, positive in the direction of rotation, and $w$ being the vertical velocity, positive in the same sense as the Northern Galactic Pole. We calculate $u$, $v$, $w$ using a combination of the measured proper motions (in the tangential velocities) and radial velocities, 
\begin{align}
\begin{split}
\label{eqn:uvw}
u = (v_r &+ k_1)\cos b \cos(\phi + \ell) - (v_\ell + k_2)\sin(\phi + \ell) \\ &- (v_b + k_3) \sin b \cos(\phi + \ell) \\ 
v = (v_r &+ k_1)\cos b \sin(\phi + \ell) + (v_\ell + k_2)\cos(\phi + \ell) \\& - (v_b + k_3) \sin b \sin(\phi + \ell) \\ 
w = (v_r &+ k_1)\sin b + (v_b + k_3) \cos b 
\end{split}
\end{align}
in which $k_1$, $k_2$ and $k_3$ account for the Solar motion,
\begin{align}
\begin{split}
k_1 ={}&U_\odot\cos\ell \cos b + V_{g,\odot}\sin\ell \cos b + W_\odot \sin b  \\
k_2 ={}&-U_\odot\sin\ell + V_{g,\odot}\cos\ell \\
k_3 ={}&-U_\odot \cos \ell \sin b - V_{g,\odot} \sin \ell \sin b + W_\odot \cos b  
\end{split}
\end{align}
where ($U_\odot$, $V_\odot$, $W_\odot$) describe the Solar peculiar motion and $V_{g, \odot}$ is the azimuthal velocity of the Sun about the Galactic centre, given by $V_{g, \odot}$=$V_0 + V_\odot$ with $V_0$ the azimuthal velocity of the LSR. We adopt for these the values of \cite{McMillan2017}: $U_\odot=8.6\pm0.9$\,km\,s$^{-1}$, $V_{\odot}=13.9\pm1$\,km\,s$^{-1}$, $W_\odot=7.1\pm1.0$\,km\,s$^{-1}$, $V_{g,\odot}=247\pm3$\,km\,s$^{-1}$, and distance of the Sun to the Galactic centre $R_0=8.20\pm0.09$.

\section{Results}
\label{sec:results}
The data used here is available through CDS, including positions, distances, velocities etc of the final sample (see also the Appendix).

\subsection{Radial motion}
\label{sec:u-vs-RG}

\begin{figure*}
 \centering
  \includegraphics[width=\textwidth]{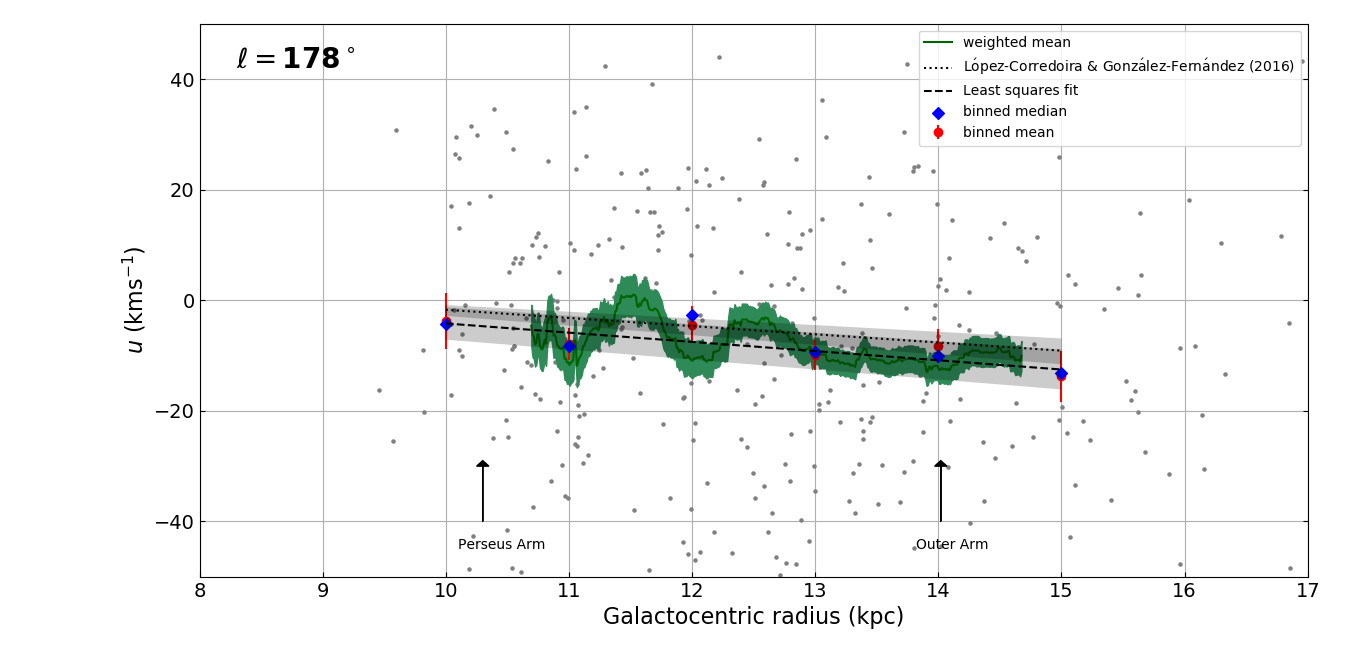}
  \includegraphics[width=\textwidth]{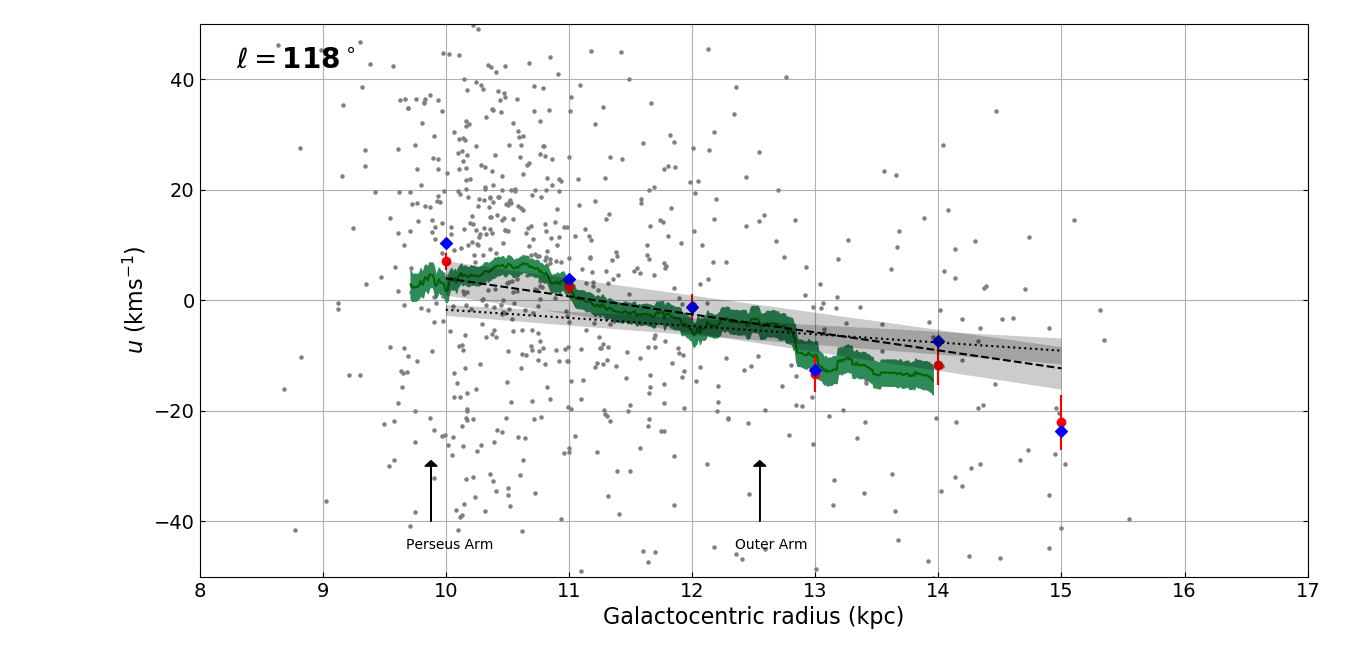}
 \caption{\it{The trend of u with $R_G$ for $\ell=178^\circ$ (top) and $\ell=118^\circ$ (bottom). The green line is the weighted mean of the grey data points, and the shaded region represents the standard error of the mean. The red points are the mean u of 1\,kpc bins, and the error bars are the standard error of the binned mean. The blue diamonds are the median u of 1\,kpc bins. The black dashed line is a weighted linear regression line fit to the grey data points. The black dotted line is the \citet{Lopez-Corredoira2016} result. Arrows indicate the approximate location of the Perseus and Outer Arms \citep{Reid2014}.}}
\label{fig:u-vs-RG}
 \end{figure*}

The trend of $u$ with Galactocentric distance, $u(R_G)$, is shown in figure \ref{fig:u-vs-RG} for $\ell=178^\circ$ (top) and $\ell=118^\circ$ (bottom). Both sightlines show an overall negative gradient in $u$, determined from the weighted linear regression line (dashed line) fit to the data points, along with some wiggles in the weighted mean trend (green line, shaded to represent standard error of the mean). This negative gradient has been measured previously. For example, \citet{Lopez-Corredoira2016} and \citet{Tian2017} have both used clump giants located near the anticentre direction to achieve this (note that their definition of $u$ is opposite in sign to ours). The \citet{Lopez-Corredoira2016} result is plotted in figure \ref{fig:u-vs-RG} as a black dotted line, and is very similar to our result in the anticentre direction. From the line fitted to our data at $\ell=178^\circ$, we find a gradient in Galactocentric radial velocity of $du/dR_G=-1.67\pm0.14$\,km\,s$^{-1}$\,kpc$^{-1}$, with a zero point at $R_G(u=0)=7.46\pm1.13$\,kpc, to be compared with \citet{Lopez-Corredoira2016} result of $du/dR_G=-1.48\pm0.35$\,km\,s$^{-1}$\,kpc$^{-1}$ and $R_G(u=0)=8.84\pm2.74$\,kpc. \citet{Tian2017} do not fit a linear trend but find the radial profile crosses $u=0$ at $R_G\sim9$\,kpc which is slightly further out than our measurement. Since the stated values depend on the adopted Solar motion which varies between studies, we have recomputed our results switching to the adopted Solar motions of these earlier works. Our results remain compatible with \citet{Lopez-Corredoira2016}, and still fall short of the $R_G\sim9$\,kpc cross-point obtained by \citet{Tian2017}.

The linear fit to the radial velocity profile at $\ell=118^\circ$ is notably different. The measured gradient is steeper than in the anticentre, and the $u=0$ crosspoint is further out: we measure $du/dR_G=-3.25\pm0.15$\,km\,s$^{-1}$\,kpc$^{-1}$ and $R_G(u=0)=11.23\pm0.71$\,kpc for the range covered. However if we move away from the idea of a linear trend and examine  the mean trend directly, we notice at $\ell=118^\circ$ the profile is almost step-like with a section of $u\sim10$\,km\,s$^{-1}$ for $R_G<11$, a section of $u\sim0$ from $11<R_G$ (kpc) $13$, and a section of $u\sim-10$\,km\,s$^{-1}$ for $R_G>13$\,kpc.

The wiggles in the running means are likely to be due to the noise level of the data - hence why it is more noticeable in the less well-sampled $\ell=178^\circ$ sightline. However, kinematic perturbations in the radial direction, for example linked to spiral arms and/or the bar, could be present. We explore possible explanations for the observed behaviour in the discussion (section \ref{sec:disc1}).

\subsection{Azimuthal motion - the rotation curve}
\label{sec:v-vs-RG}

\begin{figure*}
 \centering
  \includegraphics[width=\textwidth]{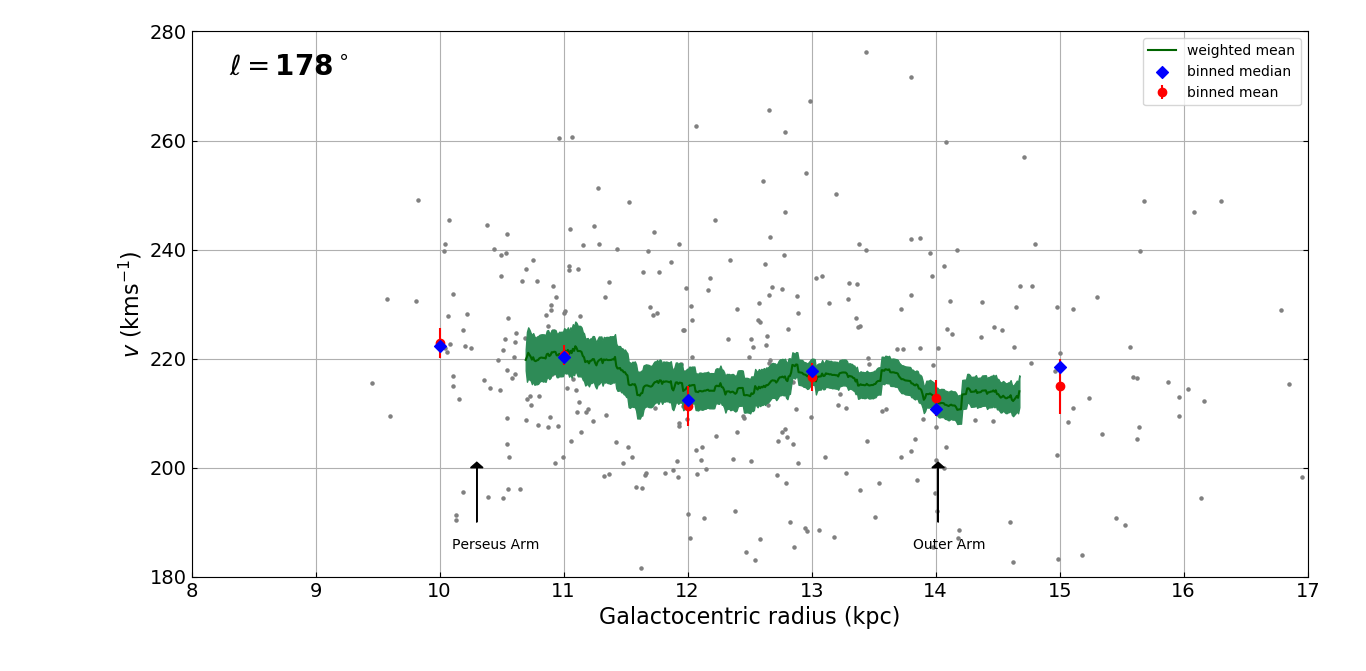}
  \includegraphics[width=\textwidth]{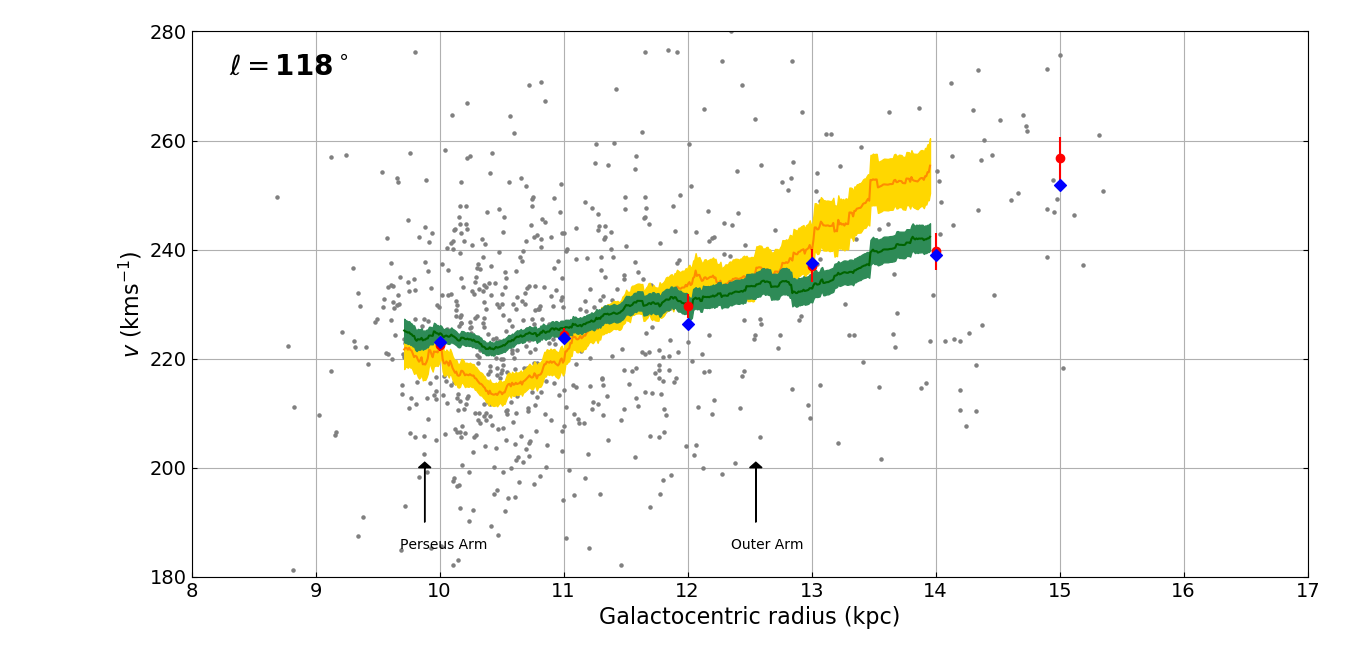}
 \caption{\it{The rotation curve for $\ell=178^\circ$ (top) and $\ell=118^\circ$ (bottom). The green line is the weighted mean of the grey data points, and the shaded region represents the standard error of the mean. The red points are the mean v of 1\,kpc bins, and the error bars are the standard error of the binned mean. The blue diamonds are the median v of 1\,kpc bins. The yellow line shows the result of H18 for comparison. Arrows indicate the approximate location of the Perseus and Outer Arms \citep{Reid2014}.}}
\label{fig:rotcurve}
 \end{figure*}

The rotation curve, $v(R_G)$, measured at $\ell=178^\circ$ is shown in the top panel of figure \ref{fig:rotcurve} (green line). It is roughly flat. The absolute value at which it lies scales directly with the assumed solar motion, ($U_\odot$, $V_{g,\odot}$, $W_\odot$)\footnote{It also scales with the assumed $R_0$, but this is much less significant than the effect from assumed Solar motion.} . For our adopted solar azimuthal velocity $V_{g,\odot}=247$\,km\,s$^{-1}$, we measure a mean rotation speed over $R_G\sim11-15$\,kpc of $\sim215$\,km\,s$^{-1}$. More generally, it is $\sim32$\,km\,s$^{-1}$ slower than the adopted rotation speed of the Sun ($v - V_{g,\odot}=215 - 247 = -32$\,km\,$^{-1}$). This is consistent with the findings of \citet{Kawata2018}, who use Gaia DR2 proper motions for a very large sample ($>10^6$) of stars located along the Galactic centre-anticentre line to determine the rotation speed. At $R_G=10-12$\,kpc, they measure the rotation speed to be $\sim31$\,km\,s$^{-1}$ slower than their assumed $V_{g,\odot}$. Our work confirms this result and almost doubles their range measured in the outer disk, extending to $R_G=15$\,kpc thanks to the greater reach of our spectro-photometric distance scale.


The rotation curve measured at $\ell=118^\circ$, shown in the bottom panel of figure \ref{fig:rotcurve}, is not the flat profile observed in the anticentre. Instead we observe a gradual increase from $\sim222$\,km\,s$^{-1}$ at 10.5\,kpc to $242$\,km\,s$^{-1}$ near 14\,kpc - that is an increase of $\sim20$\,km\,s$^{-1}$ over $R_G=10.5-14$\,kpc.

A rising rotation law has been measured before, for example by \citet{Tian2017} in the anticentre direction. It was also measured in H18, who use the same sample as here but with only the heliocentric radial velocity data available. In this situation, it was necessary to make an assumption about the behaviour of $u(R_G)$. H18 chose to treat $u(R_G)$ as averaging to zero at all distances along the pencil beam. The H18 result, recalculated using the Solar motion adopted in this paper, is plotted on the $\ell=118^\circ$ panel in figure \ref{fig:rotcurve} for comparison (yellow line). We see that the new result from this paper, able to take into account proper motion measurements, is slightly flatter than the previous result in H18 - it does not dip as low at $R_G\sim10.5$\,kpc or reach as high at $R_G>13$\,kpc. Clearly, the presence of a significant radial velocity term that does not average to zero has an impact on the rotation curve deduced from observed stellar motions.

\citet{Huang2016} also measured a sharply rising rotation curve between $R_G=11-15$\,kpc that is very similar to that of H18 by using clump giants sampled over a broad fan of outer disk longitudes. However, unlike H18 they did not make the assumption of zero radial motion, and instead treated a longitude-averaged $u(R_G)$ as a free parameter in their kinematic model of the Galaxy. The trend they find is much weaker than the $u(R_G)$ trend we find here at $\ell=118^\circ$.

\subsection{Vertical motion}
\label{sec:w-vs-RG}

\begin{figure*}
 \centering
  \includegraphics[width=\textwidth]{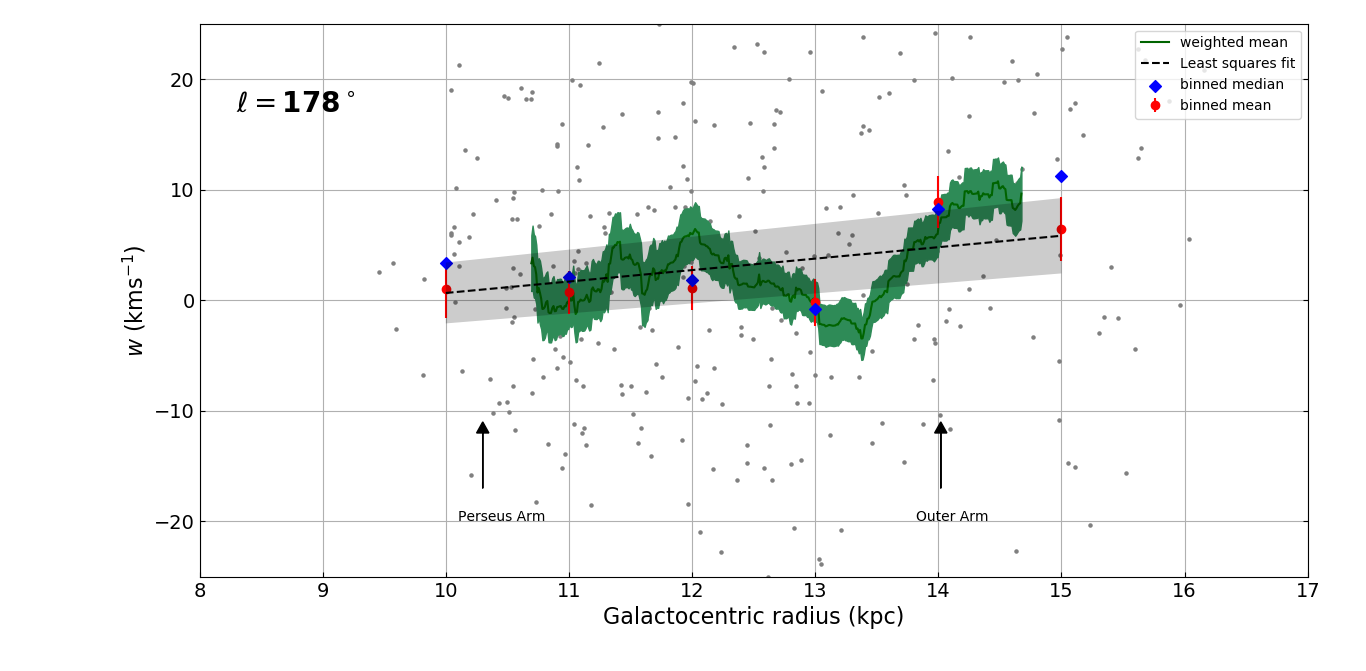}
  \includegraphics[width=\textwidth]{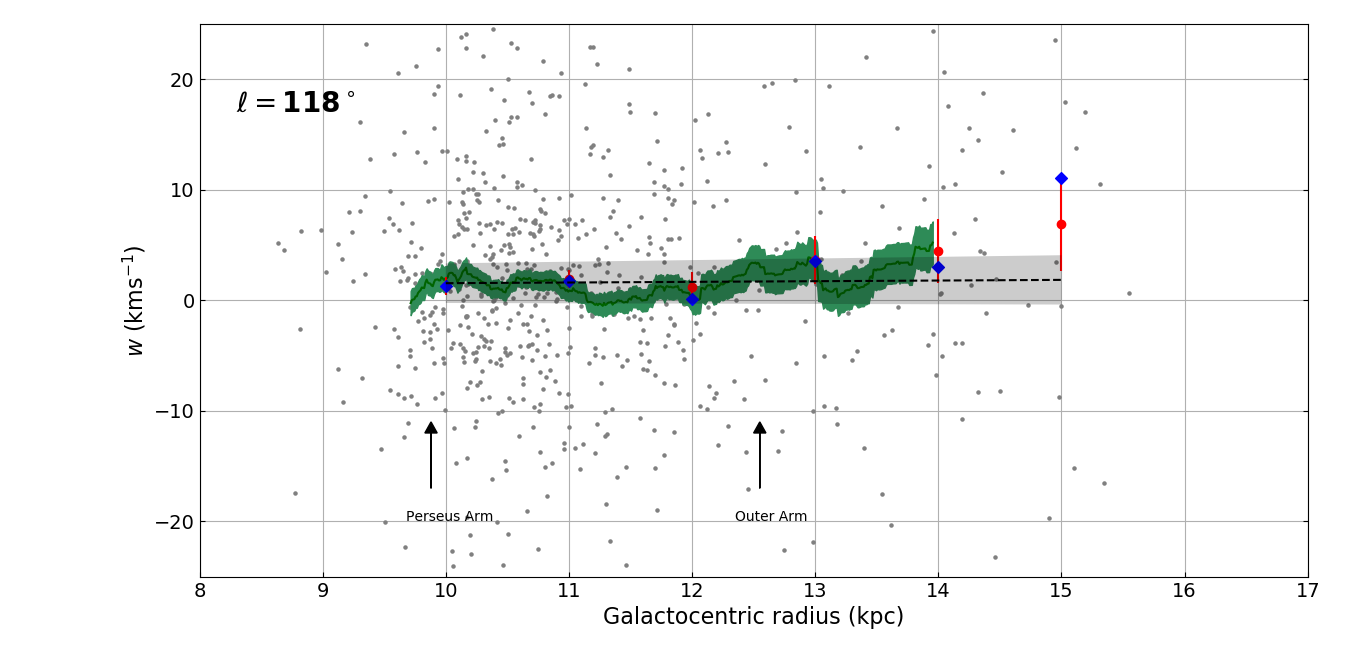}
 \caption{\it{The trend of w with $R_G$ for $\ell=178^\circ$ (top) and $\ell=118^\circ$ (bottom). The green line is the weighted mean of the grey data points, and the shaded region represents the standard error of the mean. The red points are the mean w of 1\,kpc bins, and the error bars are the standard error of the binned mean. The blue diamonds are the median w of 1\,kpc bins. The black dashed line is a weighted linear regression line fit to the grey data points.  Arrows indicate the approximate location of the Perseus and Outer Arms \citep{Reid2014}. Note the vertical scale is $2\times$ more sensitive than in the equivalent $u$ and $v$ plots.}}
\label{fig:w-vs-RG}
 \end{figure*}

Figure \ref{fig:w-vs-RG} shows the vertical velocities as a function of Galactocentric distance, $w(R_G)$, at $\ell=178^\circ$ (top panel) and $\ell=118^\circ$ (bottom panel). 

In the case of $\ell=118^\circ$, the trend is roughly flat. The weighted linear regression line (black dashed) has a slope of $0.06\pm0.09$\,km\,s$^{-1}$\,kpc$^{-1}$ - consistent with zero gradient. On average the vertical velocity is slightly positive at $\sim2$\,km\,s$^{-1}$, although this scales with the assumed $W_\odot$. 

At $\ell=178^\circ$ we observe something different. Firstly, the weighted linear regression (black dashed line) returns a slope of $dw/dR_G=1.03\pm0.13$\,km\,s$^{-1}$\,kpc$^{-1}$. Secondly, the weighted mean (green line) indicates the trend is also oscillating. Whilst this wiggling may be in-part due to low number statistics, a similar effect has been noted in previous studies: most recently, \citet{Kawata2018} use Gaia DR2 proper motions of stars along the Galactic centre-anticentre line out to $R_G=12$\,kpc to find a positive gradient of vertical velocity with Galactocentric radius and they also observe oscillations around this gradient. The results of \citet{SchonrichDehnen2018}, based on the Gaia-TGAS data set, exhibited this behaviour also. We discuss possible explanations of these perturbations in section \ref{sec:disc1}. 

\subsection{A and F star comparison - radial motion, asymmetric drift and vertex deviation}
\label{sec:u-v-w}
In order to determine if there are any intrinsic differences in the kinematics of the A and F stars measured, we examine the $u-v$ plane. Figure \ref{fig:178-u-v} shows the $\ell=178^\circ$ objects in the $u-v$ plane, split into two distance bins: an inner region $10<R_G$ (kpc) $\leq13$, and an outer region $13<R_G$ (kpc) $\leq16$. There are a total of 235 stars (113 A and 122 F) in the inner region, and 117 stars (93 A and 24 F) in the outer region. Similarly, figure \ref{fig:118-u-v} shows the $\ell=118^\circ$ objects in the $u-v$ plane, in distance bins of $9<R_G$ (kpc) $\leq11$, and $11<R_G$ (kpc) $\leq14$. The $\ell=118^\circ$ distance bins are better sampled than at $\ell=178^\circ$, with 420 stars (225 A and 195 F) in the inner region and 257 stars (185 A and 72 F) in the outer region. We compare the kinematics of the two stellar types in each distance bin, keeping in mind that the inner bin of each sightline has the largest sample size and most comparable number of A and F stars, and the $\ell=118^\circ$ sightline in general has the larger number of stars, providing the more robust statistics. 

\subsubsection{Radial motion}
The median $u$ values (vertical dashed lines) are consistent within 1$\sigma$ for the different stellar types in both distance bins at $\ell=178^\circ$. However in the better sampled $\ell=118^\circ$ distance bins, the $u$ value for A stars is positively offset from the F stars by $\sim10$\,km\,s$^{-1}$. It is not obvious (to us) why these population samples should exhibit this difference in both the distance regions. Nevertheless, both stellar groups exhibit a positive (inward) radial motion inside $R_G=11$\,kpc, while outwards motion is the norm beyond this radius.


\subsubsection{Asymmetric drift}

A slight lag in azimuthal velocity of F stars compared to A stars would not be a surprise, since they are slightly older and hence have been subject to larger kinematic scatter and have had more time to build up asymmetric drift. For our sample, the median $v$ values (horizontal dashed lines) are consistent to within 1$\sigma$ in all panels in both sightlines, except for the $\ell=178^\circ$ outer region where the F stars lag the A stars by $12$\,km\,s$^{-1}$. This is the least populated distance bin with only 24 F stars, and hence the result is accompanied by significant error. Hence, our results indicate the difference in asymmetric drift between A and F stars is negligible.

Asymmetric drift, $v_a$, in young stars is small, but the exact magnitude expected for A and F stars is not well known. \citet{DehnenBinney1998} use Hipparcos data to study the kinematics of main sequence stars as a function of $B-V$ colour. Figure 10.12 of \citet{BinneyMerrifield1998} shows the $v_a$ values of their sample. For late A stars with $B-V\sim0.2$, they find $v_a=4-5$\,km\,s, and for early F stars with $B-V\sim0.4$, they find $v_a=5-6$\,km\,s$^{-1}$. \citet{Robin2017} model the asymmetric drift as a function of $R_G$ and $Z$, and similarly find for stars younger than 1\,Gyr the asymmetric drift is $\lesssim3$\,km\,s$^{-1}$ in the plane of the disk, increasing by just $\sim1$\,km\,s$^{-1}$ for stars aged $1-2$\,Gyr. \citet{Kawata2018b} apply an axisymmetric disk model to 218 Galactic Cepheids - young objects like those in our sample - and find negligible asymmetric drift of $0.28\pm0.2$\,km\,s$^{-1}$ at $R_0$. Clearly the consistent theme from previous work is that A/F star asymmetric drift is small and hence will not significantly affect our measured rotation curve.

\begin{figure*}
 \centering
  \includegraphics[width=0.49\textwidth]{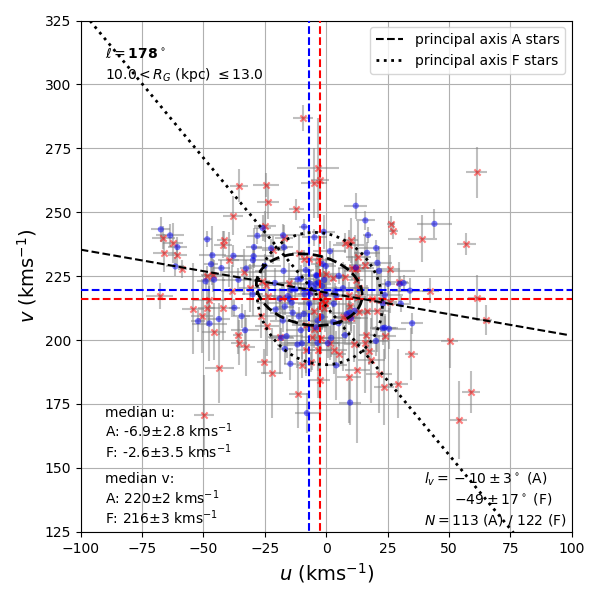}
  \includegraphics[width=0.49\textwidth]{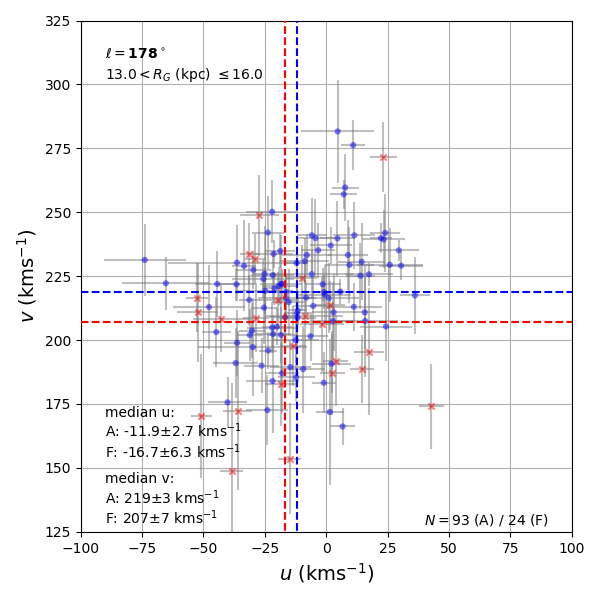}
 \caption{\it{The $\ell=178^\circ$ objects in the u-v plane split into two distance ranges: an inner region $10<R_G$ (kpc) $\leq13$ (left), and an outer region $13<R_G$ (kpc) $\leq16$ (right). Blue circles represent A stars and red crosses represent F stars. The blue (red) dashed lines show the median u and v values for A (F) stars. The black dashed (dotted) lines show the velocity ellipsoid and its major axis, defining the vertex deviation, for the A (F) stars. }}
\label{fig:178-u-v}
 \end{figure*}

\begin{figure*}
 \centering
  \includegraphics[width=0.49\textwidth]{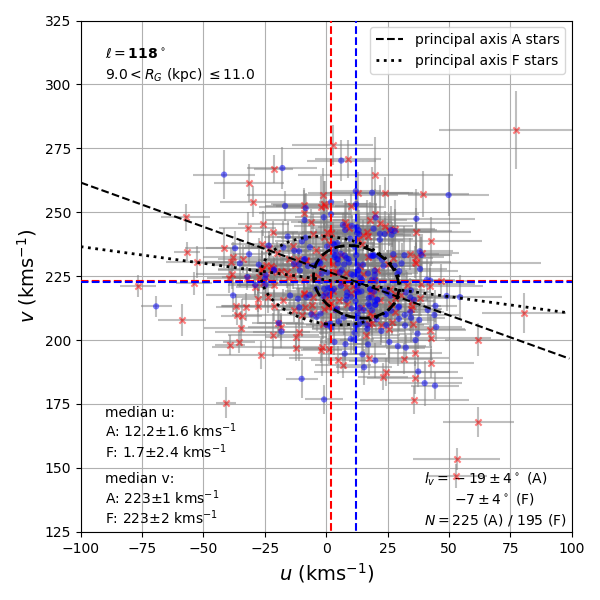}
  \includegraphics[width=0.49\textwidth]{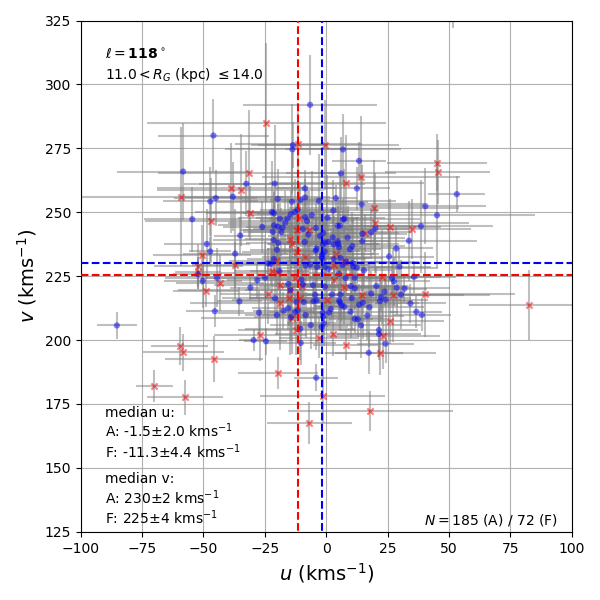}
 \caption{\it{The $\ell=118^\circ$ objects in the u-v plane, split into two distance ranges: an inner region $9<R_G$ (kpc) $\leq11$ (left), and an outer region $11<R_G$ (kpc) $ \leq14$ (right).  Blue circles represent A stars and red crosses represent F stars. The blue (red) dashed lines show the median u and v values for A (F) stars. The black dashed (dotted) lines show the velocity ellipsoid and its major axis, defining the vertex deviation, for the A (F) stars.}}
\label{fig:118-u-v}
 \end{figure*}

\subsubsection{Vertex deviation}
For the distance ranges with comparable numbers of A and F stars, we calculate the vertex deviation for the two stellar types. The vertex deviation, $l_v$, is the Galactic longitude at which the principal (or major) axis of the velocity ellipsoid is aligned, given by
\begin{equation}
l_v = 0.5 \tan ^{-1} \left( \frac{2\sigma_{uv}^2}{\sigma_u^2 - \sigma_v^2} \right)
\end{equation}
where $\sigma_u$ and $\sigma_v$ are the velocity dispersions in the $u$ and $v$ velocity components (see table \ref{tab:vel-1}), and $\sigma_{uv}^2=\overline{(u-\bar{u})(v - \bar{v})}$, with the superposed bar representing the average. Following \citet{Vorobyov2006} and to account for possible large deviations, we correct $l_v$ with
\begin{equation}
l_v =
\begin{cases}
l_v & \sigma_u^2 > \sigma_v^2 \\
l_v + sign(\sigma_{uv}^2)\frac{\pi}{2} & \sigma_u^2 < \sigma_v^2
\end{cases}
\end{equation}
We determine the vertex deviation and its uncertainty for the A and F samples using Monte Carlo simulations, drawing each $u$ and $v$ value from a gaussian distribution with spread determined by their individual errors, and calculating $l_v$ with this drawn sample. We do this 1000 times, and take the mean and standard deviation of the resulting distribution as the measured vertex deviation and its corresponding uncertainty. The left panels of figures \ref{fig:178-u-v} and \ref{fig:118-u-v} show the velocity ellipsoid and principal axis drawn for A stars and F stars. The vertex deviations and velocity dispersions perpendicular to the principal axis, $\sigma_1$, are detailed in table \ref{tab:vel-1}, along with the ratio of the minor to major axis dispersions, $\sigma_2/\sigma_1$. 

At $\ell=118^\circ$ the vertex deviations are $-19\pm4^\circ$ for the A stars and $-7\pm4^\circ$ for the F stars. The ratio of velocity dispersion of the minor and major axes is $0.77$ for the A stars and $0.60$ for the F stars, implying the ellipsoids are well-defined. At $\ell=178^\circ$ we see a large difference in the vertex deviation of the F stars compared to the A stars. For the A stars, $l_v$ is $-10\pm3^\circ$, whereas for the F stars it is $-49\pm17^\circ$. However the F star vertex deviation has significant uncertainty and the ellipsoid is less elliptical with $\sigma_2 = 0.91\sigma_1$.

The magnitudes of our measured vertex deviations in the better-sampled $\ell=118^\circ$ sightline are comparable with those from \citet{DehnenBinney1998}, who use Hipparcos data of the Solar Neighbourhood to determine the vertex deviation of late F stars (with $B-V\sim0.5$) to be $\sim10^\circ$, rising to  $\sim30^\circ$ for early A stars (with $B-V\sim0.05$). Note, however, that the sign of our result is opposite. \citet{Roca-Fabrega2014} show that the sign of the vertex deviation changes when moving either across a spiral arm, or across a main resonance of the spiral arms (e.g. the Corotation radius or outer Lindblad resonance, OLR). In the two distance ranges over which $l_v$ is calculated, the majority of the stars lie beyond where a spiral arm is believed to be located \citep{Reid2014}. The sign of our measured vertex deviation is potentially a response to this. To test this, a larger sample size on the near-side (far-side) of the Perseus (Outer) Arm is required. 

\begin{table*}
\begin{tabular}{cc|cc|ccl}
\multicolumn{1}{c}{\textbf{Sightline \& distance}}                                           & \textbf{A/F} & \textbf{$\sigma_u$ (kms$^{-1}$)} & \textbf{$\sigma_v/\sigma_u$} & \textbf{$\sigma_1$ (kms$^{-1}$)} & \textbf{$\sigma_2/\sigma_1$} & \textbf{$l_v$ (deg)} \\ \hline
\multirow{2}{*}{\begin{tabular}[c]{@{}c@{}}178$^\circ$\\ $10<R_G$ (kpc) $\leq13$\end{tabular}} & A            & $21.0\pm1.4$                     & $0.74\pm0.06$                & $21.7\pm1.8$                     & $0.63\pm0.06$                & $-10\pm3$            \\
                                                                                              & F            & $29.3\pm3.4$                     & $0.73\pm0.10$                & $26.8\pm2.5$                     & $0.91\pm0.11$                & $-49\pm17$            \\ \hline
\multirow{2}{*}{\begin{tabular}[c]{@{}c@{}}118$^\circ$\\ $9<R_G$ (kpc) $\leq11$\end{tabular}}  & A            & $16.1\pm0.8$                     & $0.95\pm0.06$                & $17.8\pm0.8$                     & $0.77\pm0.04$                & $-19\pm4$            \\
                                                                                              & F            & $29.0\pm2.2$                     & $0.60\pm 0.05$               & $28.5\pm1.7$                     & $0.60\pm0.05$                & $-7\pm4$           
\end{tabular}
\caption{\it{Velocity dispersions and vertex deviations for the A and F stars at $R_G=10-13$\,kpc at $\ell=178^\circ$ and $R_G=9-11$\,kpc at $\ell=118^\circ$. $\sigma_u$ and $\sigma_v$ are dispersions along the Galactocentric radial and azimuthal directions, and $\sigma_1$ and $\sigma_2$ are dispersions along the principal axes of the velocity ellipsoid.}}
\label{tab:vel-1}
\end{table*}



\section{Discussion}
\label{sec:disc1}

The perturbation of stellar kinematics is an effect of a non-axisymmetric Galactic potential. It is well known that the Milky Way hosts a central bar and spiral structure, and hence the potential departs from axisymmetry. It is then to be expected that we find structure in the observed velocity profiles. In order to fully explain our results, the effects of all non-axisymmetric perturbers should be considered simultaneously, but as stated in \citet{Minchev2010}, the individual effects of these perturbers do not add up linearly. However, we cautiously proceed to examine our results and discuss possible explanations for our findings.

\subsection{The central bar}
The central bar perturbs the velocity field near its natural resonances \citep{Contopoulos1980, Dehnen2000, Muhlbauer2003, BinneyTremaine2008}. The general picture is as follows. Inside the OLR, stellar orbits become elongated perpendicular to the major axis of the bar. Outside the OLR, they become elongated parallel to the bar. The elongation of these orbits result in perturbations of the Galactocentric radial velocity that depend on the radius of the orbit and the angle relative to the bar's major axis, $\phi_b$. The perturbations are strongest close to the radius of the OLR, and are modulated by $\sin2\phi_b$ resulting in the radial gradient being strongest at e.g. $\phi_b=45^\circ$. The bar pattern speed, and consequently the location of the OLR, continues to be debated \citep[e.g.][]{Portail2015, Sormani2015}, with estimates ranging between $R_G\sim6$ and $\sim12$\,kpc.  

\citet{Muhlbauer2003} (hereafter MD03) modelled the effect of the Galactic bar on the outer disk, locating the OLR at $0.92R_0$. They found that the magnitude of the measured radial perturbation in the dominant $m=2$ mode is dependent on the radial velocity dispersion of the population, and hence stellar age. The signature is more pronounced for younger stars thanks to less radial smoothing (see MD03 figure 4). 
Across the OLR, a sharp step-like feature is predicted in the perturbation as the sign of $u$ switches from negative to positive with a total amplitude of $10-15$\,km\,s$^{-1}$ for younger populations. With increasing $R_G$ beyond the OLR, this perturbation dies away until, in MD03's model, it is negligible at $\sim1.4 R_0$ (or $R_G \sim 11.5$~kpc). The magnitude of perturbation scales with the strength of the bar potential.  Perturbations from higher order modes such as $m=4$ are also expected, but their magnitude of effect is considerably smaller -- as recently confirmed by \citet{Hunt2018} who explored a range of bar models.

At $\ell=118^\circ$, $\phi_b$ ranges from $\sim0-15^\circ$ (for a bar oriented at $\phi=30^\circ$, see figure \ref{fig:sketch}),whereas $\phi_b\sim30^\circ$ at $\ell=178^\circ$. Since bar perturbation is strongest at $\phi_b=45^\circ$, a weaker signal is expected at $\ell=118^\circ$ than at $\ell=178^\circ$. However, comparing our $u(R_G)$ profiles, the overall amplitude of change in $u$ at $\ell=118^\circ$ is greater than at $\ell=178^\circ$. 
Additionally, there is no clear evidence of the sign switch toward more positive $u$ signalling the OLR, and so the data do not inform us about the location of the OLR.  We have to conclude bar perturbation is not the dominant factor shaping the observed radial velocity profiles.

\subsection{Spiral structure}
Spiral arms also give rise to non-axisymmetric perturbation. The scale of perturbation expected depends on the model adopted for the creation of the spiral arms. For example, \citet{Monari2016} find Galactocentric radial velocity perturbations of order $\pm5$\,km\,s$^{-1}$ within and between the arms, by simulating the effect of a spiral potential on the Milky Way thin stellar disk. Observationally, \citet{Grosbol2018} support this by using B and A type stars in the Galactic centre direction to measure radial perturbations of $3-4$\,km\,s$^{-1}$, and although they focus on density wave theory their results can not exclude a transient perturbation. 

Examining our results at $\ell=178^\circ$, the wiggles in the mean trend of $u$ in figure \ref{fig:u-vs-RG} do not correlate with the location of spiral arms (black arrows), and seem to have a wavelength too short to link to spiral arm perturbations expected from spiral density wave theory. These bumps could be noise due to low-number statistics. At $\ell=118^\circ$ we see a small bump in the trend at $10-11$\,kpc, slightly further out than the Perseus Arm, and also possibly at $12-13$\,kpc, close to the Outer Arm, but it is difficult to determine if these are real features or if again, they are noise. This was already evident in the results of H18.

Studies favouring the transient winding arm view predict a change in the velocity field across spiral arms. The general picture of a spiral arm in its mid-life phase is that, on the trailing side of the arm, stars rotate more slowly and move radially outwards, whereas stars on the leading side rotate faster and move radially inwards \citep{Grand2012, Kawata2014}. \citet{Grand2016} consider the transient winding arm model and find the radial perturbation of young stars ($<3$\,Gyr) to be considerably stronger at up to $\pm20$\,km\,s$^{-1}$ across the loci of the arm. They find the azimuthal perturbation to be of order $\sim10$\,km\,s$^{-1}$. \citet{Baba2018} (hereafter B18) use Cepheids with Gaia DR1 data to confirm a velocity field that changes on crossing the Perseus Arm. However, they measure the trailing side to be rotating faster and moving inward toward the Galactic centre compared to the leading side - i.e of opposite sense to a transient arm in its mid-life phase. They attribute this pattern to the arm being in the disruption phase.

In order to test whether there is a difference in the velocity field on either side of an arm, as predicted from transient winding arm theory, we follow the concept of B18 and compare median $u$ and $v$ values within $0.2-1.5$\,kpc of the locus of the Outer Arm \citep[as determined by][]{Reid2014}. We restrict our analysis to the Outer Arm since we have comparable sample sizes for the two (leading \& trailing) sides in both sightlines. 
Our results are summarised in table \ref{tab:baba}. In our sightlines the trailing side of the arm has a smaller $R_G$ than the leading. 

The median $u$ and $v$ values are consistent within 1$\sigma$ errors across the Outer Arm at $\ell=178^\circ$, going against expectations of transient winding arm theory. However, there is a significant difference in median $u$ values and $v$ values on either side of the Arm at $\ell=118^\circ$. The leading side is rotating faster than the trailing side. At the same time, the arm is migrating radially outwards on the whole, but the leading side is moving faster and hence the Arm appears to be expanding in this sightline. This points to a combination of $u$ and $v$ perturbations that do not fit consistently with the mid-life stage of the transient winding arm model.

\begin{table}
\begin{tabular}{cc|ccc}
 &  & u & v & N \\ \hline
\multirow{2}{*}{$\ell=178^\circ$} & Trailing & $-9.3\pm3.2$ & $217\pm3$ & 94 \\
 & Leading & $-11.8\pm4.7$ & $213\pm5$ & 35 \\ \hline
\multirow{2}{*}{$\ell=118^\circ$} & Trailing & $-0.6\pm2.3$ & $227\pm2$ & 169 \\
 & Leading & $-11.7\pm3.5$ & $243\pm3$ & 61
\end{tabular}
\caption{\it{Median u and v values (km\,s$^{-1}$) for the trailing side and leading side of the Outer Arm at $\ell=178^\circ$ (top rows) and $\ell=118^\circ$ (bottom rows) . The number of objects in each subsample is also shown.}}
\label{tab:baba}
\end{table}

\subsection{Satellites and the warp}

Perturbations of vertical velocities can arise due to the passage of a satellite galaxy through the Galactic disk. \citet{Gomez2013} simulate the vertical density waves induced by the Sagittarius dwarf galaxy, and find vertical velocity perturbations of up to $8$\,km\,s$^{-1}$ in the outer disk. \citet{Antoja2018} explore the vertical phase-space of more than 6 million Gaia DR2 stars and reveal a spiral-like distribution which could be caused by the latest passage of the Sagittarius dwarf galaxy, between 300 and 900\,Myr ago.  Additionally, large-scale systematic vertical velocities are expected due to the warp in the anticentre region. \citet{Poggio2018} confirm this using Gaia DR2 kinematics of both upper main sequence stars and giants, measuring an increase in vertical velocity of $5-6$\,km\,s$^{-1}$ over $R_G=8-14$\,kpc (i.e $dw/dR_G\sim1$\,km\,s$^{-1}$\,kpc$^{-1}$).

As noted in section \ref{sec:w-vs-RG}, we observe a statistically significant positive gradient of $dw/dR_G=1.03\pm0.13$\,km\,s$^{-1}$\,kpc$^{-1}$ at $\ell=178^\circ$. This is in good agreement with the gradient measured in \citet{Poggio2018} due to the warp. The kinematic response to the warp is expected to be strongest in the anticentre direction, and hence it is unsurprising that we measure a steeper gradient at $\ell=178^\circ$ than at $\ell=118^\circ$. The oscillations observed in vertical velocity at $\ell=178^\circ$ may be a kinematic signature from a satellite crossing the plane of the Milky Way, such as the Sagittarius dwarf galaxy as discussed in \citet{Gomez2013}. It is expected that perturbations of this type present as mainly radially-dependent perturbations throughout the disk \citep[see figure 5 of][]{Gomez2013}. However, we do not observe obviously correlated perturbations in the $\ell=118^\circ$ sightline, which somewhat complicates the picture.

\section{Conclusions}
\label{sec:conclusions}
We have reused the sample of A and F stars from H18 with their radial velocities and spectro-photometric distances, and crossmatched with Gaia DR2 to bring in proper motions in order to obtain full space motions. Our spectro-photometric distance scale reaches almost twice as far as those that can be inferred from Gaia DR2 parallaxes. For example, \citet{Kawata2018} use Gaia DR2 parallaxes to sample out to $R_G=12$\,kpc - a distance of under 4\,kpc from the Sun in the anticentre direction, whereas our spectro-photometric scale extends to $R_G\sim15$\,kpc - almost 7\,kpc from the Sun in the anticentre direction. We have examined the profile of radial, azimuthal and vertical velocities with Galactocentric radius. Our main results are:
\begin{itemize}

\item We measure the rotation curve in the anticentre direction to be roughly flat (figure \ref{fig:rotcurve}), at a value that is $32$\,km\,s$^{-1}$ slower than the speed of the Sun. This confirms the recent work of \citet{Kawata2018}, but with the use of our spectro-photometric distances we extend this result to just short of $R_G=15$\,kpc. 
\item The rotation curve at $\ell=118^\circ$ is different from that at $\ell=178^\circ$ - it rises outwards from $R_G=10.5$\,kpc. We measure an increase of circular speed of roughly $+20$\,km\,s$^{-1}$ in the range $10.5<R_G$ (kpc) $<14$, from $222$\,km\,s$^{-1}$ to $242$\,km\,s$^{-1}$. Additionally, we note the significance of the $u$ behaviour in the determination of the rotation curve. In H18, our previous study, only radial velocity data were available and hence an assumption about the mean $u$ behaviour as a function of $R_G$ was required. This resulted in a rotation curve with a steeper incline than here. By making use of the full space motions, as we do here, a rising rotation curve is still apparent but moderated compared to that of H18.
\item The gradient in the Galactocentric radial velocity profile ($du/dR_G$) is surprisingly steep at $\ell=118^\circ$ compared to the more gentle gradient found at $\ell=178^\circ$ that has also been reported in previous studies (see figure \ref{fig:u-vs-RG}). This rules out the central bar as a dominant perturber shaping the $u(R_G)$ profile. We also find no consistent interpretation of our  Outer Arm velocities in terms of the transient winding arm model. Additionally, we find that radial expansion at large radii is observed in both sightlines. It is apparent that in order to fully explain our results, a model encompassing all perturbers simultaneously may be required.  
\item The vertical velocity profile in the anticentre direction has a positive gradient of $dw/dR_G=1.03\pm0.13$\,km\,s$^{-1}$ which could be a signature of the warp in the outer disc. This profile also shows some oscillation. Previous studies have also noted this. This is not replicated in the $\ell=118^\circ$ direction. 
\item We have separated and compared the median in-plane velocities and vertex deviations of the A and F stars. At $\ell=118^\circ$ the radial velocity of the A stars is larger than that of the F stars by $\sim10$\,km\,s$^{-1}$, however at $\ell=178^\circ$ there is no perceptible difference between the two types of star. Why the A and F stars at $\ell=118^\circ$ should exhibit so marked a difference out to large $R_G$ is unclear. The azimuthal velocities are broadly consistent between the two populations. This implies that asymmetric drift is comparable for A and F stars.
\end{itemize}

For this work we are priviledged to have full space motions of our sample and hence are able to reframe them as Galactocentric velocities with no prior assumptions. By examining only two sightlines in the outer disk, we have demonstrated the significant kinematic variations resulting from the real non-axisymmetric Galactic disk. The need grows to acknowledge departures from a simple uniform rotation law for the Galactic disk. A rotation law that deviates from the often-used flat law has strong consequences for kinematic distance determinations. For example, an object that is really at a distance of $8$\,kpc at $\ell = 118^\circ$ would deliver a radial velocity that would return a distance of $\sim6$\,kpc if the assumed rotation law was the same flat law as applies at $\ell =178^\circ$. This is an error in the region of $30\%$.


There is a history of reported velocity anomalies in a longitude range that includes the $\ell=118^\circ$ sightline. \citet{BrandBlitz1993}, for instance, presented a map of observed  H\,\textsc{ii} region radial velocities projected onto the Galactic plane, revealing surprisingly little variation of radial velocity over the range $110^\circ<\ell<140^\circ$. \citet{Russeil2003} reported a departure of $-21\pm10.3$\,km\,s$^{-1}$ from mean circular speed related to the Perseus Arm over $90^\circ<\ell <150^\circ$. CO and H\,\textsc{i} data, as presented in \citet{Reid2016spiral}, favour differing velocities in this longitude range, with  H\,\textsc{i} prefering typically more negative radial velocities than the CO, to the tune of $\sim10-20$\,km\,s$^{-1}$.  Further studies like the one here will begin to fill out the picture of how exactly such anomalies arise and set contraints on their origin. 


The opportunities for the kind of work presented here will increase as the next generation of massively-multiplexed spectrographs come into use. As an example, the WHT (William Herschel Telescope) Enhanced Area Velocity Explorer \citep[WEAVE, ][]{Dalton2016} will complete the kinematics of the northern hemisphere stars, capturing much of the outer Galactic disk. Here, and in H18, we have established a method that can turn results on younger earlier-type stars from such a facility into new and distinctive insights into disk structure.

\section{Acknowledgements}
 We thank Walter Dehnen for helpful comments regarding perturbations from the central bar. We are also grateful to the anonymous referee for useful comments which helped to improve the paper. AH acknowledges support from the UK's Science and Technology Facilities Council (STFC), grant no. ST/K502029/1. JED and MM acknowledge funding via STFC grants ST/J001333/1 and ST/M001008/1. This work has made use of data from the European Space Agency (ESA) mission
{\it Gaia} (\url{https://www.cosmos.esa.int/gaia}), processed by the {\it Gaia}
Data Processing and Analysis Consortium (DPAC,
\url{https://www.cosmos.esa.int/web/gaia/dpac/consortium}). Funding for the DPAC
has been provided by national institutions, in particular the institutions
participating in the {\it Gaia} Multilateral Agreement.



\bibliographystyle{mn2e}
\bibliography{write_up_bib5}

\appendix

\section{Table of data}
The data used in this paper is available through CDS. Table \ref{tab:appendix} details the presented columns.

\begin{table*}
\begin{tabular}{llll}
\textbf{Column} & \textbf{Label} & \textbf{Units}                & \textbf{Description}                                                           \\ \hline
1               & ID             & ---                           & Target ID                                                                      \\
2               & RAJ2000          & deg                           & Right ascension (J2000.0)                                                      \\
3               & DEJ2000          & deg                           & Declination (J2000.0)                                                          \\
4               & GLON           & deg                           & Galactic longitude                                                             \\
5               & GLAT           & deg                           & Galactic latitude                                                              \\
6               & HRV            & km\,s$^{-1}$   & Heliocentric radial velocity                                                   \\
7               &  e\_HRV     & km\,s$^{-1}$   & Radial velocity negative error                                                 \\
8               &  E\_HRV      & km\,s$^{-1}$   & Radial velocity positive error                                                 \\
9               & Dist           & kpc                           & Heliocentric distance                                                          \\
10              & e\_Dist      & kpc                           & Heliocentric distance error                                                        \\
11              & RG             & kpc                           & Galactocentric distance                                                        \\
12              & e\_RG        & kpc                           & Galactocentric distance error                                                  \\
13              & SourceID       & ---                           & Gaia Source ID                                                                 \\
14              & pmlcosb        & mas\,yr$^{-1}$ & Proper motion in Galactic longitude direction (multiplied by $\cos$(b) factor) \\
15              &  e\_pmlcosb   & mas\,yr$^{-1}$ & Error of proper motion in Galactic longitude direction                         \\
16              & pmb            & mas\,yr$^{-1}$ & Proper motion in Galactic latitude direction                                   \\
17              &  e\_pmb       & mas\,yr$^{-1}$ & Error of proper motion in Galactic latitude direction                          \\
18              & u              & km\,s$^{-1}$   & Galactocentric radial velocity                                                 \\
19              &  e\_u         & km\,s$^{-1}$   & Galactocentric radial velocity error                                           \\
20              & v              & km\,s$^{-1}$   & Galactocentric azimuthal velocity                                              \\
21              &  e\_v         & km\,s$^{-1}$   & Galactocentric azimuthal velocity error                                        \\
22              & w              & km\,s$^{-1}$   & Galactocentric vertical velocity                                               \\
23              &  e\_w         & km\,s$^{-1}$   & Galactocentric vertical velocity error                                        
\end{tabular}
\label{tab:appendix}
\caption{\it{The columns presented in the table of data, available through CDS.}}
\end{table*}

\label{lastpage}

\end{document}